\documentclass[journal]{IEEEtran}
\usepackage{amsmath}
\usepackage{amssymb}
\usepackage{amsfonts}
\usepackage{graphicx}
\usepackage{epsfig}
\usepackage{subfigure}
\usepackage{psfrag}
\usepackage{color}

\begin{document}

\title{Dynamic Resource Allocation in Cognitive Radio Networks: A Convex Optimization Perspective}

\author{Rui Zhang, Ying-Chang Liang, and Shuguang Cui \vspace{-0.05in}}

\maketitle

\IEEEpeerreviewmaketitle

\setlength{\baselineskip}{1.0\baselineskip}
\newtheorem{claim}{Claim}
\newtheorem{guess}{Conjecture}
\newtheorem{definition}{Definition}
\newtheorem{fact}{Fact}
\newtheorem{assumption}{Assumption}
\newtheorem{theorem}{\underline{Theorem}}[section]
\newtheorem{lemma}{\underline{Lemma}}[section]
\newtheorem{ctheorem}{Corrected Theorem}
\newtheorem{corollary}{\underline{Corollary}}[section]
\newtheorem{proposition}{Proposition}
\newtheorem{example}{\underline{Example}}[section]
\newtheorem{remark}{\underline{Remark}}[section]
\newtheorem{problem}{\underline{Problem}}[section]
\newtheorem{algorithm}{\underline{Algorithm}}[section]
\newcommand{\mv}[1]{\mbox{\boldmath{$ #1 $}}}

\begin{abstract}
This article provides an overview of the state-of-art results on
communication resource allocation over space, time, and frequency
for emerging cognitive radio (CR) wireless networks. Focusing on the
interference-power/interference-temperature (IT) constraint approach
for CRs to protect primary radio transmissions, many new and
challenging problems regarding the design of CR systems are
formulated, and some of the corresponding solutions are shown to be
obtainable by restructuring some classic results known for
traditional (non-CR) wireless networks. It is demonstrated that
convex optimization plays an essential role in solving these
problems, in a both rigorous and efficient way. Promising research
directions on interference management for CR and other related
multiuser communication systems are discussed.
\end{abstract}

\section{Introduction}

In recent years, \emph{cognitive radio} (CR) networks, where CRs or
the so-called secondary users (SUs) communicate over certain
bandwidth originally allocated to a primary network, have drawn
great research interests in the academic, industrial, and regulation
communities. Accordingly, there is now a rapidly growing awareness
that CR technology will play an essential role in enabling {\it
dynamic spectrum access} for the next generation wireless
communications, which could hopefully resolve the spectrum scarcity
vs. under-utilization dilemma caused by the current static spectrum
management polices. Specifically, the users in the primary network,
or the so-called primary users (PUs), could be licensed users, who
have the absolute right to access their spectrum bands, and yet
would be willing to share the spectrum with the unlicensed SUs.
Alternatively, both the PUs and SUs could equally coexist in an
unlicensed band, where the PUs are regarded as existing active
communication links while the SUs are new links to be added. A
unique feature of CRs is that they are able to identify and acquire
useful environmental information (cognition) across the primary and
secondary networks, and thereby adapt their transmit strategies to
achieve the best performance while maintaining a required quality of
service (QoS) for each coexisting active primary link. Depending on
the type of cognitive knowledge collected (e.g., on/off statuses of
primary links, PU messages, interference power levels at PU
receivers, or primary link performance margins) and the
primary/secondary network models of interests (e.g.,
infrastructure-based vs. ad hoc), many new and challenging problems
on the design of CR networks can be formulated, as will be reviewed
in this article.

To date, quite a few operation models have been proposed for CRs;
however, there is no consensus yet on the terminology used for the
associated definitions \cite{Mitola,Zhao07,Goldsmith08}. Generally
speaking, there are two basic operation models for CRs:
Opportunistic Spectrum Access (OSA) vs. Spectrum Sharing (SS). In
the OSA model, the SUs are allowed to transmit over the band of
interest when all the PUs are not transmitting at this band. One
essential enabling technique for OSA-based CRs is {\it spectrum
sensing}, where the CRs individually or collaboratively detect
active PU transmissions over the band, and decide to transmit if the
sensing results indicate that all the PU transmitters are inactive
at this band with a high probability. Spectrum sensing is now a very
active area for research; the interested readers may refer to, e.g.,
\cite{Quan08a,Liang08,Li08,Zeng} for an overview of the state-of-art
results in this area. As a counterpart, the SS model allows the SUs
to transmit simultaneously with PUs at the same band even if they
are active, provided that the SUs know how to control their
resultant interference at the PU receivers such that the performance
degradation of each active primary link is within a tolerable
margin. Thus, OSA and SS can be regarded as the {\it
primary-transmitter-centric} and {\it primary-receiver-centric}
dynamic spectrum access techniques, respectively. Consequently,
there will be an inevitable debate on which operation model, OSA or
SS, is better to deploy CRs in practical systems; however, a
rigorous comparative study for these two models, in terms of
spectrum efficiency and implementation complexity tradeoffs, is
still open. Generally speaking, SS utilizes the spectrum more
efficiently than OSA, since the former supports concurrent PU and SU
transmissions over the same band while the latter only allows
orthogonal transmissions between them. Moreover, the
receiver-centric approach for SS is more effective for CRs to manage
the interference to the PU links than the transmitter-centric
approach for OSA.

Hence, the SS model for CRs will be focused in this article. It is
worth noting that the optimal design approach for SS-based CR
networks should treat all coexisting PU and SU links as a giant
interference network and jointly optimize their transmissions to
maximize the SU network throughput with a prescribed PU network
throughput guarantee. From this viewpoint, recent advances in
network information theory \cite{Kumar00} have provided promising
guidelines to approach the fundamental limits of such networks.
However, from a practical viewpoint, the centralized design approach
for PU and SU networks is not desirable, since PU and SU systems
usually belong to different operators and thus it is difficult, if
not infeasible, for them to cooperate. Consequently, a {\it
decentralized} design approach is more favorable, where the PU
network is designed without the awareness of the existence of the SU
network, while the SU network is designed with only partial
knowledge (cognition) of the PU network.

Following this (simplified) decentralized approach, there are
furthermore two design paradigms proposed for SS-based CRs. One is
based on the ``cognitive relay'' concept \cite{Tarokh06}, where the
SU transmitter allocates only part of its power to deliver the SU
messages, and uses the remaining power to relay the PU messages so
as to compensate for the additional SU interference experienced at
the PU receiver. However, this technique requires non-causal
knowledge of the PU messages at the SU transmitter, which may be
difficult to realize in practice. In contrast, a more feasible SS
design for the SU to protect the PU is to impose a constraint on the
maximum SU interference power at the PU receiver, also known as the
``interference temperature (IT)'' constraint \cite{Gastpar07}, by
assuming that the SU-to-PU channels are either perfectly known at
the SU transmitters, or can be practically estimated.

In this article, we will focus our study on the IT-based SS model
for CRs, namely the IT-SS, as it is a more feasible approach
compared with other existing ones. In a wireless communication
environment, channels are usually subject to space-time-frequency
variation due to multipath propagation, mobility, and
location-dependent shadowing. Thus, {\it dynamic resource
allocation} (DRA) becomes an essential technique for CRs to
optimally deploy their transmit strategies to maximize the secondary
network throughput, where the transmit power, bit-rate, bandwidth,
and antenna beam should be dynamically allocated based upon the
available channel state information (CSI) of the primary and
secondary networks. In particular, this article will focus on DRA
problem formulations unique to CR systems under the IT-SS model, and
the associated solutions that are non-obvious in comparison with
existing results \cite{Luo06a} known for the traditional (non-CR)
wireless networks. More importantly, we will emphasize the key role
of various {\it convex optimization} techniques in solving these
problems.

The remainder of this article is organized as follows. Section
\ref{sec:system model} presents different models of the SU network
coexisting with the PU network, and various forms of transmit power
and interference power (IT) constraints over the SU transmissions.
Section \ref{sec:CR MIMO} is devoted to the spatial-domain transmit
optimization at the SUs for different SU networks subject to
transmit and interference power constraints. Section
\ref{sec:spacetime DRA} extends the results to the more general case
of joint space-time-frequency transmit optimization of the SUs, and
addresses the important issue on how to optimally set the IT
thresholds in CR systems to achieve the best spectrum sharing
performance. Finally, conclusions are drawn and future research
directions are discussed in Section \ref{sec:conclusion}.

{\it Notation:} Lower-case and upper-case bold letters denote
vectors and matrices, respectively. ${\rm Rank} (\cdot)$, ${\rm
Tr}(\cdot)$, $|\cdot|$, $(\cdot)^{-1}$, $(\cdot)^H$, and
$(\cdot)^{1/2}$ denote the rank of a matrix, trace, determinant,
inverse, Hermitian transpose, and square-root, respectively. ${\bf
I}$ and ${\bf 0}$ denote an identity matrix and an all-zero matrix,
respectively. ${\rm Diag}({\bf a})$ denotes a diagonal matrix with
diagonal elements given in ${\bf a}$. ${\rm E}(\cdot)$ denotes the
statistical expectation. A circularly symmetric complex Gaussian
(CSCG) distributed random vector with zero mean and covariance
matrix ${\bf S}$ is denoted by $\mathcal{CN}({\bf 0},{\bf S})$.
$\mathbb{C}^{m \times n}$ denotes the space of $m\times n$ complex
matrices. $\|\cdot\|$ denotes the $2$-norm of a complex vector.
${\rm Re}(\cdot)$ and ${\rm Im}(\cdot)$ denote the real and
imaginary parts of a complex number, respectively. The base of the
logarithm function $\log(\cdot)$ is $2$ by default.

\section{CR Network Models} \label{sec:system model}

\begin{figure}
\begin{center}
\scalebox{0.6}{\includegraphics*[28pt,542pt][400pt,774pt]{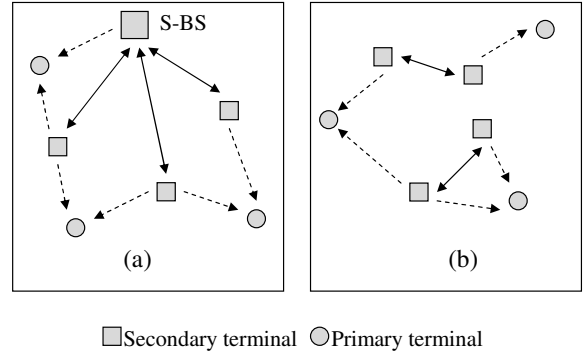}}
\end{center}\vspace{-0.1in}
\caption{CR networks: (a) infrastructure-based; (b) ad hoc.}
\label{fig:system model}\vspace{-0.1in}
\end{figure}

We consider two general types of CR networks, which are of both
theoretical and practical interests: One is {\it
infrastructure-based}, as shown in Fig.~\ref{fig:system model}(a),
where multiple secondary terminals communicate with a common
secondary node referred to as the secondary base station (S-BS); the
other is {\it ad hoc}, as shown in Fig.~\ref{fig:system model}(b),
which consists of multiple distributed secondary links. In both
types of CR networks, there are coexisting primary terminals
operating in the same spectrum band. For the IT-SS model of CRs, the
exact operation model of the primary network is not important to our
study, provided that all the secondary terminals satisfy the
prescribed IT constraints to protect the primary terminals. Without
loss of generality, we assume that there are $K$ secondary links and
$J$ primary terminals in each type of the CR networks.

Consider first the infrastructure-based secondary/CR network with
the S-BS coordinating all the CR transmissions, which usually
corresponds to one particular cell in a CR cellular network. The
uplink transmissions from the SUs to the S-BS are usually modeled by
a multiple-access channel (MAC), while the downlink transmissions
from the S-BS to different SUs are modeled by a broadcast channel
(BC). For the MAC, the equivalent baseband transmission can be
represented as
\begin{align}\label{eq:MAC}
{\bf y}=\sum_{k=1}^K{\bf H}_k{\bf x}_k+{\bf z}
\end{align}
where ${\bf y}\in\mathbb{C}^{M \times 1}$ denotes the received
signal at the S-BS, with $M$ denoting the number of antennas at
S-BS; ${\bf H}_k \in\mathbb{C}^{M \times N_k}$ denotes the channel
from the $k$th SU to S-BS, $k=1,\cdots,K$, with $N_k$ denoting the
number of antennas at the $k$th SU; ${\bf x}_k\in \mathbb{C}^{N_k
\times 1}$ denotes the transmitted signal of the $k$th SU; and ${\bf
z}\in\mathbb{C}^{M \times 1}$ denotes the noise received at S-BS. We
assume that ${\bf x}_k$'s are independent over $k$.

Similarly, the BC can be represented as
\begin{align}\label{eq:BC}
{\bf y}_k={\bf H}^H_k{\bf x}+{\bf z}_k, ~~ k=1,\cdots,K
\end{align}
where ${\bf y}_k\in\mathbb{C}^{N_k \times 1}$ denotes the received
signal at the $k$th SU; for convenience, we have used the Hermitian
transposed uplink channel matrix for the corresponding downlink
channel matrix, i.e., ${\bf H}^H_k$ denotes the channel from the
S-BS to the $k$th SU; ${\bf x}\in \mathbb{C}^{M \times 1}$ denotes
the transmitted signal from S-BS; and ${\bf z}_k\in\mathbb{C}^{N_k
\times 1}$ denotes the receiver noise of the $k$th SU. In the case
that ${\bf x}$ carries information common to all SUs, the associated
downlink transmission is usually called {\it multicast}, while if
${\bf x}$ carries independent information for different SUs, it is
called {\it unicast}.

Next, consider the ad hoc secondary/CR network, which is usually
modeled as an interference channel (IC). For convenience, we assume
that for the $k$th secondary link, $k=1,\cdots,K$, the transmitter
is denoted as SU-TX$_k$ and the receiver is denoted as SU-RX$_k$,
although in general a secondary terminal can be both a transmitter
and a receiver. The baseband transmission of the IC can be
represented as
\begin{align}\label{eq:IC}
{\bf \tilde{y}}_k={\bf H}_{kk}{\bf \tilde{x}}_k+\sum_{i=1,i\neq
k}^{K}{\bf H}_{ik}{\bf \tilde{x}}_i +{\bf \tilde{z}}_k, ~~
k=1,\cdots,K
\end{align}
where ${\bf \tilde{y}}_k\in\mathbb{C}^{B_k \times 1}$ denotes the
received signal at SU-RX$_k$, with $B_k$ denoting the number of
receiving antennas; ${\bf \tilde{x}}_k\in \mathbb{C}^{A_k \times 1}$
denotes the transmitted signal of SU-TX$_k$, with $A_k$ denoting the
number of transmitting antennas; ${\bf H}_{kk} \in\mathbb{C}^{B_k
\times A_k}$ denotes the direct-link channel from SU-TX$_k$ to
SU-RX$_k$, while ${\bf H}_{ik} \in\mathbb{C}^{B_k \times A_i}$
denotes the cross-link channel from SU-TX$_i$ to SU-RX$_k$, $i\neq
k$;  and ${\bf \tilde{z}}_k\in\mathbb{C}^{B_k \times 1}$ denotes the
noise at SU-RX$_k$. It is assumed that ${\bf \tilde{x}}_k$'s are
independent over $k$.

Furthermore, we assume that the $j$th PU, $j=1,\cdots,J$, in each
type of the CR networks is equipped with $D_j$ antennas, $D_j\geq
1$. We then use  ${\bf G}_{kj}\in\mathbb{C}^{D_j \times N_k}$ to
denote the channel from the $k$th SU to the $j$th PU in the CR MAC,
${\bf F}_{j}\in\mathbb{C}^{D_j \times M}$ to denote the channel from
S-BS to the $j$th PU in the CR BC, and ${\bf
E}_{kj}\in\mathbb{C}^{D_j \times A_k}$ to denote the channel from
SU-TX$_k$ to the $j$th PU in the CR IC. Moreover, the receiving
terminals in the secondary networks may experience interference from
active primary transmitters. For simplicity, we assume that such
interference is treated as additional noise at the secondary
receivers, and the total noise at each secondary receiving terminal
is distributed as a CSCG random vector with zero mean and the
identity covariance matrix.

Note that the (spatial) channels defined in the above CR network
models are assumed constant for a fixed transmit dimension such as
one time-block in a time-division-multiple-access (TDMA) system or
one frequency-bin in an orthogonal-frequency-division-multiplexing
(OFDM) system. In a wireless environment, these channels usually
change over time and/or frequency dimensions as governed by an
underlying joint stochastic process. As such, DRA becomes relevant
to schedule SUs into different transmit dimensions based on their
CSI. In general, the secondary transmitting terminals need to satisfy
two types of power constraints for DRA: One is due to their own
transmit power budgets; the other is to limit their resulting
interference level at each PU to be below a prescribed threshold.
These constraints can be applied over each fixed dimension as {\it
peak} power constraints, or over multiple dimensions as {\it
average} power constraints. Without loss of generality, we consider
DRA for the secondary network over $L$ transmit dimensions with
different channel realizations, with $L\geq 1$. In total, four different
types of power constraints can be defined for the secondary network.
By taking the CR MAC as an example (similarly as for the CR BC/IC),
we have
\begin{itemize}
\item {\bf Peak transmit power constraint (PTPC)}:
\begin{align}\label{eq:PTPC}
{\rm Tr} ({\bf S}_k[l]) \leq P_k
\end{align}
where ${\bf S}_k[l]$ denotes the transmit covariance matrix for the
$l$th transmit dimension of the $k$th SU, $l\in\{1,\cdots,L\},
k\in\{1,\cdots,K\}$; $P_k$ denotes the $k$th SU's peak power
constraint that applies to each of the $L$ transmit dimensions.

\item {\bf Average transmit power constraint (ATPC)}:
\begin{align}\label{eq:ATPC}
\frac{1}{L}\sum_{l=1}^L{\rm Tr} ({\bf S}_k[l]) \leq \bar{P}_k
\end{align}
where $\bar{P}_k$ denotes the $k$th SU's average transmit power
constraint over the $L$ transmit dimensions.

\item {\bf Peak interference power constraint (PIPC)}:
\begin{align}\label{eq:PIPC}
\sum_{k=1}^{K}{\rm Tr} \left({\bf G}_{kj}[l]{\bf S}_k[l]{\bf
G}^H_{kj}[l]\right) \leq \Gamma_j
\end{align}
where ${\bf G}_{kj}[l]$ denotes the realization of channel ${\bf
G}_{kj}$ for a given $l$; and $\Gamma_j$ denotes the peak
interference power constraint for protecting the $j$th PU,
$j\in\{1,\cdots,J\}$, which limits the total interference power
caused by all the $K$ SUs across all the receiving antennas of the
$j$th PU, for each of the $L$ transmit dimensions.

\item {\bf Average interference power constraint (AIPC)}:
\begin{align}\label{eq:AIPC}
\frac{1}{L}\sum_{l=1}^L\sum_{k=1}^{K}{\rm Tr} \left({\bf
G}_{kj}[l]{\bf S}_k[l]{\bf G}^H_{kj}[l]\right) \leq \bar{\Gamma}_j
\end{align}
where $\bar{\Gamma}_j$ denotes the average interference power
constraint for the $j$th PU to limit the total interference power
from the $K$ SUs, which is averaged over the $L$ transmit
dimensions.
\end{itemize}

Note that DRA for traditional (non-CR) wireless networks under PTPC
and/or ATPC has been thoroughly studied in the literature
\cite{Shamai98}, while the study of DRA subject to PIPC and/or AIPC
as well as their various combinations with PTPC/ATPC is unique to CR
networks and is relatively new. In order to gain more insights into
the optimal DRA designs for CR networks, we will first study the
case of a single transmit dimension ($L=1$) with PTPCs and PIPCs by
focusing on the spatial-domain transmit optimization for
multi-antenna CRs in Section \ref{sec:CR MIMO}, and then study the
general case of $L>1$ for joint space-time-frequency DRA in CR
networks under ATPCs and AIPCs in Section \ref{sec:spacetime DRA}.

\section{Cognitive Beamforming Optimization} \label{sec:CR MIMO}

In this section, we consider the case of $L=1$, where the DRA for CR
networks reduces to the spatial-domain transmit optimization under
PTPCs and PIPCs to maximize the CR network throughput. We term this
practice as {\it cognitive beamforming}. In order to investigate the
fundamental performance limits of cognitive beamforming, we study
the optimal designs with the availability of perfect knowledge on
all the channels in the SU networks, and those from all the
secondary transmit terminals to PUs. For convenience, we drop the
dimension index $l$ for the rest of this section given $L=1$.

First, it is worth noting that the PIPC given in (\ref{eq:PIPC}) can
be unified with the PTPC given in (\ref{eq:PTPC}) into a form of
{\it generalized linear transmit covariance constraint} (GLTCC):
\begin{align}\label{eq:GLTCC}
\sum_{i=1}^K{\rm Tr}({\bf W}_i{\bf S}_i)\leq w
\end{align}
where ${\bf W}_i$'s and $w$ are constants. For example, with each
PIPC given in (\ref{eq:PIPC}), ${\bf W}_i={\bf G}_{ij}^H{\bf
G}_{ij}, \forall i$, and $w=\Gamma_j$, while for each PTPC given in
(\ref{eq:PTPC}), ${\bf W}_i={\bf I}$ if $i=k$ and ${\bf 0}$
otherwise, with $w= P_k$. Previous studies on transmit optimization
for multi-antenna or multiple-input multiple-output (MIMO) systems
have mostly adopted some special forms of GLTCC such as the user
individual power constraints and sum-power constraint. However, it
remains unclear whether such existing solutions are applicable to the
general form of GLTCC, which is crucial to the problem of CR MIMO
transmit optimization with the newly added PIPCs. In the following,
we provide an overview of the state-of-art solutions for this
problem under different CR network models, while the developed
solutions also apply to the case with the general form of GLTCCs as
in (\ref{eq:GLTCC}). From a convex optimization perspective, we next
divide our discussions into two parts, which deal with the cases of
convex and non-convex problem formulations, respectively.

\subsection{Convex Problem Formulation} First, consider the case
where the associated optimization problem in a traditional MIMO
system without PIPC is {\it convex}. In such cases, since the extra
PIPCs are linear over the SU transmit covariance matrices, the
resulting transmit covariance optimization problem for CR systems
remains convex; and thus, it can be efficiently solved by standard
convex optimization techniques.

{\bf CR Point-to-Point MIMO Channel:} We elaborate this case by
first considering the CR point-to-point MIMO channel, which can be
treated as the special case with only one active SU link in the MAC,
BC, or IC based CR network. Without loss of generality, we will use
the notations developed for the CR MAC with $K=1$ in the following
discussions. Specifically, the optimal transmit covariance to
achieve the CR point-to-point MIMO channel capacity under both the
PTPC and PIPCs can be obtained from the following problem
\cite{Zhang08a}:
\begin{align}
\mathop{{\rm Max.}}_{\bf S} & ~~ \log\left|{\bf I}+{\bf H}{\bf S}
{\bf H}^H\right| ~~~~~~~~~~~~~~~~~~~~~~\rm(P1)\nonumber
\\ {\rm s.~t.} & ~~ {\rm Tr}({\bf S})\leq P \nonumber \\
&~~ {\rm Tr}\left( {\bf G}_j{\bf S}{\bf G}_j^H\right)\leq \Gamma_j,
~ j=1,\cdots,J \nonumber
\\ &~~ {\bf S}\succeq {\bf 0} \nonumber
\end{align}
where for conciseness we have removed the SU index $k$ in the symbol
notations since $K=1$, while ${\bf S}\succeq {\bf 0}$ means that
${\bf S}$ is a positive semi-definite matrix \cite{Boydbook}.

We see that (P1) is a convex optimization problem since its
objective function is concave over ${\bf S}$ and its constraints
define a convex set over ${\bf S}$. Thus, (P1) can be efficiently
solved by, e.g., the interior point method \cite{Boydbook}. In the
special case of CR multiple-input single-output (MISO) channel,
i.e., ${\bf H}$ degrades to a row-vector denoted by ${\bf
h}\in\mathbb{C}^{1 \times N}$, it can be shown by exploiting the
Karush-Kuhn-Tucker (KKT) optimality conditions of (P1) that transmit
beamforming is capacity optimal, i.e., ${\rm Rank}({\bf S})=1$
\cite{Zhang08a}. Thus, without loss of generality we could write
${\bf S}={\bf v}{\bf v}^H$, where ${\bf v}\in\mathbb{C}^{N \times
1}$ denotes the precoding vector. Accordingly, (P1) for the special
case of MISO CR channel is simplified as (P1-S) \cite{Zhang08a}:
\begin{align}
\mathop{{\rm Max.}}_{\bf v} & ~~ \|{\bf hv}\| \nonumber
\\ {\rm s.~t.} & ~~ \|{\bf v}\|^2\leq P \nonumber \\
&~~ \|{\bf G}_j{\bf v}\|^2 \leq \Gamma_j, ~ j=1,\cdots,J, \nonumber
\end{align}
which is non-convex due to the non-concavity of its objective
function. However, by observing the fact that if ${\bf v}$ is the
solution of (P1-S), so is $e^{j\theta}{\bf v}$ for any arbitrary
$\theta$, we thus assume without loss of generality that ${\bf hv}$
is a real number and modify (P1-S) by rewriting its objective
function as ${\rm Re}({\bf hv})$ and adding an additional linear
constraint ${\rm Im}({\bf hv})=0$. Thereby, (P1-S) can be converted
into a second-order cone programming (SOCP) \cite{Boydbook} problem,
which is convex and thus can be efficiently solved by available
convex optimization softwares \cite{CVX}. Alternatively, (P1-S) can
be shown equivalent to its Lagrange dual problem \cite{Zhang08a},
which is a convex semi-definite programming (SDP) \cite{Boydbook}
problem and is thus efficiently solvable \cite{CVX}. For (P1-S) in
the case of one single-antenna PU, a closed-form solution for the
optimal precoding vector ${\bf v}$ was derived in \cite{Zhang08a}
via a geometric approach.

In order to reveal the structure of the optimal ${\bf S}$ for (P1),
we consider its Lagrange dual problem defined as (P1-D):
\begin{align}
\mathop{{\rm Min.}}_{\mv{\eta}\succeq{\bf 0}} ~ d(\mv{\eta})
\nonumber
\end{align}
where $\mv{\eta}=[\eta_0,\eta_1,\cdots,\eta_J]$ denotes a vector of
dual variables for (P1) with $\eta_0$ associated with the PTPC, and
$\eta_j$ associated with the $j$th PIPC, $j=1,\cdots,J$, while we
have the dual function defined as
\begin{align}\label{eq:Lagrangian P1}
d(\mv{\eta})\triangleq \max_{{\bf S}\succeq {\bf 0}} &~
\log\left|{\bf I}+{\bf H}{\bf S} {\bf H}^H\right|-\eta_0({\rm
Tr}({\bf S})-P) \nonumber \\ &~ -\sum_{j=1}^J\eta_j({\rm Tr}\left(
{\bf G}_j{\bf S}{\bf G}_j^H\right)-\Gamma_j).
\end{align}
Since (P1) is convex with Slater's condition satisfied~\cite{Boydbook}, the duality gap between the optimal values of
(P1) and (P1-D) is zero, i.e., (P1) can be solved
equivalently as (P1-D). Accordingly, an iterative algorithm can be
developed to solve (P1-D) by alternating between solving
$d(\mv{\eta})$ for a given $\mv{\eta}$ and updating $\mv{\eta}$ to
minimize $d(\mv{\eta})$. At each iteration, $\mv{\eta}$ can be
updated by a subgradient-based method such as the ellipsoid method
\cite{BGT81}, according to the subgradients of $d(\mv{\eta})$, which
can be shown equal to $P-{\rm Tr}({\bf S}^{\star})$ and $\Gamma_j-{\rm
Tr}\left( {\bf G}_j{\bf S}^{\star}{\bf G}_j^H\right)$ for $\eta_0$
and $\eta_j, j\neq 0$, respectively, where ${\bf S}^{\star}$ denotes
the optimal ${\bf S}$ to obtain $d(\mv{\eta})$ for a given
$\mv{\eta}$. From (\ref{eq:Lagrangian P1}), it follows that ${\bf
S}^{\star}$ is the solution of the following equivalent problem (by
discarding irrelevant constant terms):
\begin{align}\label{eq:max P1}
\max_{{\bf S}\succeq {\bf 0}}~\log\left|{\bf I}+{\bf H}{\bf S} {\bf
H}^H\right|-{\rm Tr}({\bf T}{\bf S})
\end{align}
where ${\bf T}=\eta_0{\bf I}+\sum_{j=1}^J\eta_j({\bf G}_j^H{\bf
G}_j)$ is a constant matrix for a given $\mv{\eta}$. In order to
solve Problem (\ref{eq:max P1}), we introduce an auxiliary variable:
$\hat{{\bf S}}={\bf T}^{1/2}{\bf S}{\bf T}^{1/2}$. Problem
(\ref{eq:max P1}) is then re-expressed in terms of $\hat{{\bf S}}$
as
\begin{align}\label{eq:max P1 new}
\max_{\hat{{\bf S}}\succeq {\bf 0}}~\log\left|{\bf I}+{\bf H}{\bf
T}^{-1/2}\hat{{\bf S}}{\bf T}^{-1/2}{\bf H}^H\right|-{\rm
Tr}(\hat{{\bf S}}).
\end{align}
The above problem can be shown equivalent to the standard
point-to-point MIMO channel capacity optimization problem subject to
a single sum-power constraint \cite{Cover}, and its solution can be
expressed as $\hat{{\bf S}}^{\star}={\bf V}{\bf \Sigma}{\bf V}^H$,
where ${\bf V}$ is obtained from the singular-value decomposition
(SVD) given as follows: ${\bf H}{\bf T}^{-1/2}={\bf U}{\bf
\Theta}{\bf V}^H$, with ${\bf \Theta}={\rm
Diag}([\theta_1,\ldots,\theta_T])$ and $T=\min(M,N)$, while ${\bf
\Sigma}= {\rm Diag}([\sigma_1,\ldots,\sigma_T])$ follows the
standard water-filling solution \cite{Cover}:
$\sigma_i=(1/\ln2-1/\theta_i^2)^+, i=1,\ldots,T$, with
$(\cdot)^+\triangleq\max(0,\cdot)$. Thus, the solution of Problem
(\ref{eq:max P1}) for a given $\mv{\eta}$ can be expressed as ${\bf
S}^{\star}={\bf T}^{-1/2}{\bf V}{\bf \Sigma}{\bf V}^H{\bf
T}^{-1/2}$.

Next, we present a heuristic method for solving (P1), which leads to
a suboptimal solution in general and could serve as a benchmark to
evaluate the effectiveness of the above two approaches based on
convex optimization. To gain some intuitions for this method, we
first take a look at two special cases of (P1). For the first case,
supposing that all the PIPCs are inactive (e.g., by setting
$\Gamma_j=\infty, \forall j$) and thus can be removed, (P1) reduces
to the standard MIMO channel capacity optimization problem under the
PTPC only, for which the optimal solution of ${\bf S}$ is known to
be derivable from the SVD of ${\bf H}$ \cite{Cover}. For the second
case, assuming that $\Gamma_j=0, \forall j$, the solution for (P1)
is then obtained by the ``zero-forcing (ZF)'' algorithm
\cite{Spencer04}, which first projects ${\bf H}$ into the space
orthogonal to all ${\bf G}_j$'s, and then designs the optimal ${\bf
S}$ based on the SVD of the projected channel.  Note that the
(non-trivial) ZF-based solution exists only when $N> \sum_{j=1}^J
D_j$. From the above two special cases, we observe that as
$\Gamma_j$'s decrease, the optimal ${\bf S}$ should evolve along
with a sequence of subspaces of ${\bf H}$ with decreasing dimensions
as a result of keeping certain orthogonality to ${\bf G}_j$'s, which
motivates a new design method for cognitive beamforming, named as
{\it partial channel projection} \cite{Zhang08a}. Specifically, let
${\bf \bar{G}}_j={\bf G}_j/\Gamma_j, \forall j$. Then, define ${\bf
\bar{G}}\triangleq[{\bf \bar{G}}_1^T,\cdots,{\bf \bar{G}}_J^T]^T$.
Denote the SVD of ${\bf \bar{G}}$ as ${\bf \bar{G}}={\bf U}_G{\bf
\Lambda}_G{\bf V}_G^H$. Without loss of generality, assume that the
singular values in ${\bf \Lambda}_G$ are arranged in a decreasing
order. Then, we propose a generalized channel projection operation:
\begin{align}
{\bf H}_{\bot}={\bf H}\left({\bf I}-{\bf V}_G^{(b)}\left({\bf
V}_G^{(b)}\right)^H\right)
\end{align}
where ${\bf V}_G^{(b)}$ consists of the first $b$ columns of ${\bf
V}_G$ corresponding to the $b$ largest singular values in ${\bf
\Lambda}_G$, $1\leq b\leq \min(N-1, \sum_{j=1}^J D_j)$. Note that
$b$ could also take a zero value for which ${\bf
V}_G^{(0)}\triangleq{\bf 0}$. Now, we are ready to present the
transmit covariance matrix for the partial projection method in the
form of its eigenvalue decomposition (EVD) as ${\bf S}={\bf
V}_{\bot}{\bf \Sigma}_{\bot}{\bf V}^H_{\bot}$, where ${\bf
V}_{\bot}$ is obtained from the SVD of the projected channel ${\bf
H}_{\bot}$, i.e., ${\bf H}_{\bot}={\bf U}_{\bot}{\bf
\Lambda}_{\bot}{\bf V}_{\bot}^H$. By substituting this new form of
${\bf S}$ into (P1), it can be shown that the problem reduces to
maximizing the sum-rate of a set of parallel channels (with channel
gains given by ${\bf \Lambda}_{\bot}$) over their power allocation
${\bf \Sigma}_{\bot}$ subject to $(J+1)$ linear power constraints,
for which the optimal power allocation can be obtained by a
generalized ``water-filling'' algorithm \cite{Zhang08a}. Note that
the partial channel projection works for any values of $N$ and
$D_j$'s.

In Fig. \ref{fig:MIMO rate comp}, we plot the achievable rate of a
CR MIMO channel under the PTPC and PIPCs with the optimal transmit
covariance solution for (P1) via the convex optimization approach, against those with suboptimal covariance solutions via the
partial channel projection method with different values of $b$. The
system parameters are given as follows: $M=N=4$, $J=2$, $D_1=D_2=1$,
and $\Gamma_1=\Gamma_2=0.1$. The SU achievable rate is plotted vs.
the SU PTPC, $P$. It is observed that the optimal covariance
solution obtained via the convex optimization approach yields
notable rate gains over suboptimal solutions via the heuristic
method, for which the optimal value of $b$ (the number of SU-to-PU
channel dimensions to be nulled) to maximize the SU achievable rate
increases with the SU PTPC.

{\bf CR MIMO-MAC}: The solutions proposed for the CR point-to-point
MIMO channel shed insights to transmit optimization for the CR
MIMO-MAC defined in (\ref{eq:MAC}) with $K>1$. Assume that in the CR
MIMO-MAC, the optimal multiuser detection is deployed at the S-BS to
successively decode different SU messages from the received
sum-signal. We then consider the problem for jointly optimizing SU
transmit covariance matrices to maximize their weighted sum-rate
subject to individual PTPCs and joint PIPCs. This problem is
referred to as weighted sum-rate maximization (WSRMax). Without loss
of generality, we assume that the given user rate weights satisfy
that $\mu_1\geq \mu_2\geq \cdots\geq \mu_K\geq 0$; thus, the optimal
decoding order of users at the S-BS to maximize the weighted
sum-rate is in accordance with the reverse user index \cite{Tse98}.
Accordingly, the WSRMax for the CR MIMO-MAC can be expressed as
\begin{align}
\mathop{{\rm Max.}}_{{\bf S}_1,\cdots,{\bf S}_K} & ~~
\sum_{k=1}^K\mu_k\log\frac{\left|{\bf I}+\sum_{i=1}^{k}{\bf H}_i{\bf
S}_i{\bf H}_i^H\right|}{\left|{\bf I}+ \sum_{i=1}^{k-1}{\bf H}_i{\bf
S}_i{\bf H}_i^H\right|}\nonumber ~~~~~~~~~\rm{(P2)}
\\ {\rm s.~t.} & ~~ {\rm Tr}({\bf S}_k)\leq P_k, ~k=1,\ldots,K \nonumber \\
&~~ \sum_{k=1}^{K}{\rm Tr} \left({\bf G}_{kj}{\bf S}_k{\bf
G}^H_{kj}\right) \leq \Gamma_j, ~ j=1,\cdots,J \nonumber
\\ &~~ {\bf S}_k\succeq {\bf 0}, ~ k=1,\cdots,K. \nonumber
\end{align}

\begin{figure}
\centering{
 \epsfxsize=3.5in
    \leavevmode{\epsfbox{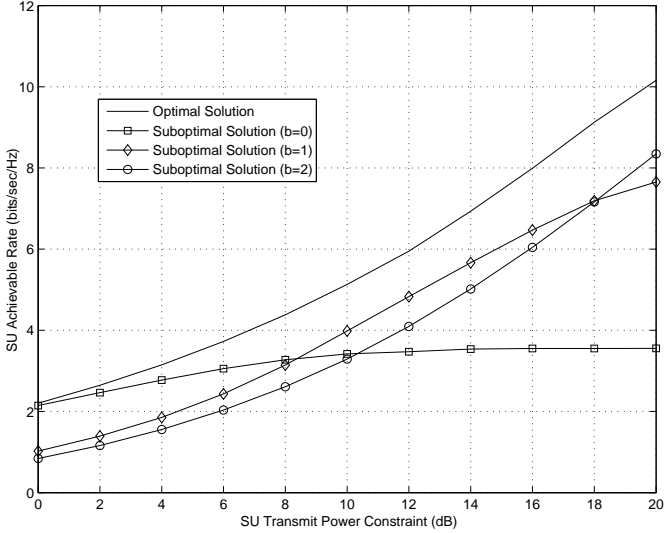}} } \vspace{-0.1in}
\caption{Comparison of the achievable rates for the CR MIMO channel
under the PTPC and PIPCs with the optimal transmit covariance
solution for (P1) via the the convex optimization approach vs.
suboptimal covariance solutions via the partial channel projection
method with different values of $b$.}\label{fig:MIMO rate
comp}\vspace{-0.2in}
\end{figure}

Reordering terms in the objective function of (P2) yields
\begin{align}
&\sum_{k=1}^{K-1}(\mu_k-\mu_{k+1})\log\left|{\bf
I}+\sum_{i=1}^{k}{\bf H}_i{\bf S}_i{\bf H}_i^H\right| \nonumber
\\ &+\mu_K \log\left|{\bf I}+\sum_{i=1}^{K}{\bf H}_i{\bf S}_i{\bf
H}_i^H\right|
\end{align}
From the above new form of the objective function, it can be verified
that (P2) is a convex optimization problem over ${\bf S}_k$'s. Thus,
similarly as for (P1), (P2) can be solved by an
interior-point-method based algorithm or an iterative algorithm via
solving the equivalent Lagrange dual problem, for which the details
are omitted here for brevity.

It is noted that (P2) is for the case with the optimal non-linear
multiuser decoder at the S-BS, while in practice the low-complexity
linear decoder is usually more preferable. The use of linear instead
of non-linear decoder at the receiver will change the user
achievable rates for the CR MIMO-MAC, thus resulting in new problem
formulations for transmit optimization. For example, in
\cite{ZhangLan08a}, the authors have considered the CR SIMO-MAC
(single-antenna for each SU transmitter) with a linear decoder at the
receiver, where the power allocation across the SUs is optimized to maximize
their signal-to-interference-plus-noise ratios (SINRs) at the
receiver subject to both transmit and interference power
constraints.

\subsection{Non-Convex Problem Formulation}

Next, we consider the case where the optimization problems in the
associated traditional (non-CR) MIMO systems are {\it non-convex}.
It thus becomes more challenging whether these non-convex problems
with the addition of convex PIPCs in the corresponding CR MIMO
systems can be efficiently solvable. In the following, we present
some promising approaches to solve these problems for the CR MIMO-BC
and MIMO-IC.

{\bf CR MIMO-BC}: First, consider the CR MIMO-BC defined in
(\ref{eq:BC}) under both the PTPC at the S-BS and $J$ PIPCs each for
one of the $J$ PUs, which can be similarly defined as for the MAC
case in (\ref{eq:PTPC}) and (\ref{eq:PIPC}), respectively. We focus
on the unicast downlink transmission for the CR BC, while for the
case of multicast, the interested readers may refer to
\cite{multicasting}. For the purpose of exposition, we consider two
commonly adopted design criteria for the traditional
multi-antenna Gaussian BC in the literature: One is for the MIMO-BC
deploying the non-linear ``dirty-paper-coding (DPC)'' at the
transmitter \cite{Shamai06}, which maximizes the weighted sum-rate
of all the users (i.e., the WSRMax problem); the other is for the
MISO-BC (single-antenna for each SU receiver) deploying only linear
encoding at the transmitter, which maximizes the minimum SINR among
all the users, referred to as ``SINR balancing''.

Specifically, the WSRMax problem for the CR MIMO-BC can be
formulated as:
\begin{align}
\mathop{{\rm Max.}}_{{\bf Q}_1,\cdots,{\bf Q}_K} & ~~
\sum_{k=1}^K\mu_k\log\frac{\left|{\bf I}+{\bf
H}_k^H\left(\sum_{i=k}^{K}{\bf Q}_i\right){\bf
H}_k\right|}{\left|{\bf I}+{\bf H}_k^H\left(\sum_{i=k+1}^{K}{\bf
Q}_i\right){\bf H}_k\right|} \nonumber ~~\rm{(P3)}
\\ {\rm s.~t.} & ~~ \sum_{k=1}^K{\rm Tr}({\bf Q}_k)\leq P \nonumber \\
&~~ {\rm Tr}\left( {\bf F}_j\left(\sum_{k=1}^K{\bf Q}_k\right){\bf
F}_j^H\right)\leq \Gamma_j, ~ j=1,\cdots,J \nonumber
\\ &~~ {\bf Q}_k\succeq {\bf 0}, ~ k=1,\cdots,K \nonumber
\end{align}
where ${\bf Q}_k\in\mathbb{C}^{M\times M}$ denotes the covariance
matrix for the transmitted signal of S-BS intended for the $k$th SU,
$k=1,\cdots,K$; $\mu_k$'s are the given user rate weights; and $P$
denotes the transmit power constraint for the S-BS. Without loss of
generality, we assume that $\mu_1\geq \mu_2\geq \cdots\geq \mu_K\geq
0$; thus, in (P3) the optimal encoding order of users for DPC to
maximize the weighted sum-rate is in accordance with the user index
\cite{Shamai06}. Note that (P3) is non-convex with or without the
PIPCs due to the fact that the objective function is non-concave
over ${\bf Q}_k$'s for $K\geq 2$. As a result, unlike (P1) for the
point-to-point CR channel, the standard Lagrange duality method
cannot be applied for this problem. For (P3) in the case without the
PIPCs, a so-called ``BC-MAC duality'' relationship was proposed in
\cite{Goldsmith03} to transform the non-convex MIMO-BC problem into
an equivalent convex MIMO-MAC problem, which is solvable by
efficient convex optimization techniques such as the interior point
method. In \cite{Yu}, another form of BC-MAC duality, the so-called
``mini-max duality'' was explored to solve the MIMO-BC problem under
a special case of GLTCC: the per-antenna transmit power constraint.
However, these existing forms of BC-MAC duality are yet unable to
handle the case with arbitrary numbers of GLTCCs, which is the case
for (P3) with both the PTPC and PIPCs.

In \cite{ZhangLan09a}, a general method was proposed to solve
various MIMO-BC optimization problems under multiple GLTCCs, thus
including the CR MIMO-BC WSRMax problem given in (P3). For this
method, the first step is to combine all $(J+1)$ power constraints
in (P3) into a single GLTCC as shown in the following optimization
problem:
\begin{align}
\mathop{{\rm Max.}}_{{\bf Q}_1,\cdots,{\bf Q}_K} & ~~
\sum_{k=1}^K\mu_k\log\frac{\left|{\bf I}+{\bf
H}_k^H\left(\sum_{i=k}^{K}{\bf Q}_i\right){\bf
H}_k\right|}{\left|{\bf I}+{\bf H}_k^H\left(\sum_{i=k+1}^{K}{\bf
Q}_i\right){\bf H}_k\right|} \nonumber
\\ {\rm s.~t.} &~~ {\rm Tr}\left( {\bf A}\sum_{k=1}^K{\bf
Q}_k\right) \leq Q \nonumber
\\ &~~ {\bf Q}_k\succeq {\bf 0}, ~ k=1,\cdots,K \label{eq:single
GLTCC BC}
\end{align}
where ${\bf A}=\lambda_0{\bf I}+\sum_{j=1}^J\lambda_j{\bf F}_j^H{\bf
F}_j$, and $Q=\lambda_0P+\sum_{j=1}^J\lambda_j\Gamma_j$ with
$\lambda_0,\lambda_1,\cdots,\lambda_J$ being non-negative constants.
For a given set of $\lambda_i$'s, $i=0,\cdots,J$, let the optimal
value of the above problem be denoted by
$F(\lambda_0,\lambda_1,\cdots,\lambda_J)$. Clearly,
$F(\lambda_0,\lambda_1,\cdots,\lambda_J)$ is an upper bound on the
optimal value of (P3) since any feasible solutions for (P3) must
satisfy the constraints of Problem (\ref{eq:single GLTCC BC}) for a
given set of $\lambda_i$'s. Interestingly, it can be shown that the
optimal value of (P3) is equal to the minimum value of function
$F(\lambda_0,\lambda_1,\cdots,\lambda_J)$ over all non-negative
$\lambda_i$'s \cite{ZhangLan09a}. Therefore, (P3) can be resolved by
iteratively solving Problem (\ref{eq:single GLTCC BC}) for a given
set of $\lambda_i$'s and updating $\lambda_i$'s towards their
optimal values to minimize function
$F(\lambda_0,\lambda_1,\cdots,\lambda_J)$. Specifically,
$\lambda_i$'s can be updated via the ellipsoid method according to
the subgradients of $F(\lambda_0,\lambda_1,\cdots,\lambda_J)$, which
can be shown \cite{ZhangLan09a} equal to $P-\sum_{k=1}^K{\rm Tr}({\bf
Q}_k^{\star})$ and $\Gamma_j-{\rm Tr}( {\bf F}_j(\sum_{k=1}^K{\bf
Q}_k^{\star}){\bf F}_j^H)$ for $\lambda_0$ and $\lambda_j$ $(j\neq
0)$, respectively, where ${\bf Q}_k^{\star}$'s are the solution of
Problem (\ref{eq:single GLTCC BC}) for the given $\lambda_k$'s.

\begin{figure}
\psfrag{A}{${\bf x}$}\psfrag{B}{${\bf H}_1^H$}\psfrag{C}{${\bf
H}_2^H$}\psfrag{D}{${\bf H}_K^H$} \psfrag{E}{${\bf
z}_1$}\psfrag{F}{${\bf z}_2$}\psfrag{G}{${\bf z}_K$}
\psfrag{H}{${\bf y}_1$}\psfrag{I}{${\bf y}_2$}\psfrag{J}{${\bf
y}_K$}\psfrag{K}{${\bf x}_1$}\psfrag{M}{${\bf x}_2$}\psfrag{O}{${\bf
x}_K$} \psfrag{L}{${\bf H}_1$}\psfrag{N}{${\bf
H}_2$}\psfrag{P}{${\bf H}_K$} \psfrag{Q}{${\bf z}$}\psfrag{R}{${\bf
y}$}
\begin{center}
\scalebox{0.7}{\includegraphics*[75pt,558pt][440pt,744pt]{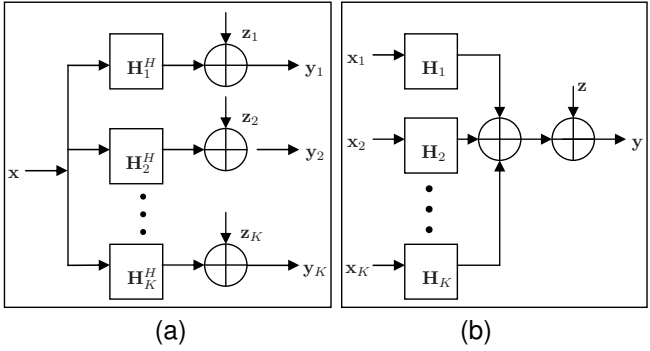}}
\end{center}
\caption{Generalized MIMO MAC-BC Duality: (a) Primal MIMO-BC channel
with downlink channels ${\bf H}_k^H$ and receiver noise vectors
${\bf z}_k\sim\mathcal{CN}({\bf 0},{\bf I})$, $k=1,\ldots,K$, and a
GLTCC: ${\rm Tr}({\bf A}\sum_{k=1}^K{\bf Q}_k) \leq Q$; (b) Dual
MIMO-MAC with uplink channels ${\bf H}_k, k=1,\ldots,K$ and receiver
noise vector ${\bf z}\sim\mathcal{CN}({\bf 0},{\bf A})$, and a
sum-power constraint: $\sum_{k=1}^{K}{\rm Tr}({\bf S}_k) \leq Q$.
The MIMO-BC and dual MIMO-MAC have the same achievable rate region
\cite{ZhangLan09a}.}\label{fig:BC MAC duality}\vspace{-0.15in}
\end{figure}

Furthermore, Problem (\ref{eq:single GLTCC BC}) with a given set of
$\lambda_k$'s can be solved by applying the {\it generalized BC-MAC
duality} proposed in \cite{ZhangLan09a}, which extends the existing
forms of BC-MAC duality \cite{Goldsmith03,Yu} to transform the
MIMO-BC problem subject to a single GLTCC as in Problem
(\ref{eq:single GLTCC BC}) to an auxiliary (dual) MIMO-MAC problem
subject to a corresponding sum-power constraint. Specifically, it is
shown in \cite{ZhangLan09a} that the MIMO-BC as in Problem
(\ref{eq:single GLTCC BC}) and the dual MIMO-MAC, as depicted in
Fig. \ref{fig:BC MAC duality}, have the same achievable rate region.
Accordingly, the optimal objective value (weighted sum-rate) of
Problem (\ref{eq:single GLTCC BC}) for the primal MIMO-BC can be
obtained as that of the following equivalent problem for the dual
MIMO-MAC:
\begin{align}
\mathop{{\rm Max.}}_{{\bf S}_1,\cdots,{\bf S}_K} & ~~
\sum_{k=1}^{K-1}(\mu_k-\mu_{k+1})\log\left|{\bf
A}+\sum_{i=1}^{k}{\bf H}_i{\bf S}_i{\bf H}_i^H\right| \nonumber
\\ &~~ +\mu_K \log\left|{\bf A}+\sum_{i=1}^{K}{\bf H}_i{\bf S}_i{\bf
H}_i^H\right|\nonumber
\\ {\rm s.~t.}
&~~ \sum_{k=1}^{K}{\rm Tr} \left({\bf S}_k\right) \leq Q \nonumber
\\ &~~ {\bf S}_k\succeq {\bf 0}, ~ k=1,\cdots,K. \label{eq:dual MAC
max}
\end{align}
Similar to (P2), the above problem is a WSRMax problem for the
MIMO-MAC subject to a single sum-power constraint, which is convex
and thus can be efficiently solvable by, e.g., the interior point
method. After solving Problem (\ref{eq:dual MAC max}), the optimal
user transmit covariance solutions for the MIMO-MAC, ${\bf
S}_k^{\star}$'s, can be transformed to the corresponding ones for
the original MIMO-BC, ${\bf Q}_k^{\star}$'s, via a MAC-BC covariance
transformation algorithm given in \cite{ZhangLan09a}. Furthermore,
it is worth noting that with $K=1$, the above method can be shown
equivalent to that developed for (P1) in the CR point-to-point MIMO
channel case based on the Lagrange duality.

Consider next the SINR balancing problem for the CR MISO-BC, which
can be expressed as:
\begin{align}
\mathop{{\rm Max.}}_{\alpha,{\bf v}_1,\cdots,{\bf v}_K} & ~~ \alpha
\nonumber ~~~~~~~~~~~~~~~~~~~~~~~~~~~~~~~~~~~\rm{(P4)}
\\ {\rm s.~t.} & ~~ \frac{\|{\bf h}^H_k{\bf v}_k\|^2}{1+\sum_{i\neq k}\|{\bf h}^H_k{\bf v}_i\|^2}\geq \alpha , ~ k=1,\cdots,K\nonumber \\
& ~~ \sum_{k=1}^K \|{\bf v}_k\|^2 \leq P \nonumber \\
&~~ \sum_{k=1}^K \|{\bf F}_j{\bf v}_k\|^2\leq \Gamma_j, ~
j=1,\cdots,J \nonumber
\end{align}
where $\alpha$ denotes an achievable SINR for all the SUs; ${\bf
v}_k\in\mathbb{C}^{M\times 1}$ denotes the precoding vector for the
transmitted signal of S-BS intended for the $k$th SU; and ${\bf
h}_k$ represents ${\bf H}_k$ for the MISO-BC case. Similarly as for
(P1-S), by treating  ${\bf h}^H_k{\bf v}_k$ on the left-hand side
(LHS) of each SINR constraint in (P4) as a positive real number
\cite{Ottersten01}, it can be shown that (P4) for a given $\alpha$
is equivalent to a SOCP feasibility problem and thus efficiently
solvable \cite{CVX}. For a given $\alpha$, if the associated SOCP
problem is feasible, we know that the optimal solution of (P4) for
$\alpha$, denoted by $\alpha^{\star}$, must satisfy
$\alpha^{\star}\geq \alpha$; otherwise, $\alpha^{\star}<\alpha$.
Based on this fact, $\alpha^{\star}$ can be found by a simple
bisection search \cite{Boydbook}; with $\alpha^{\star}$, the
corresponding optimal solution for ${\bf v}_k$'s in (P4) can also be
obtained. The above technique has also been applied in
\cite{Wiesel06} for (P4) without the PIPCs.

The SINR balancing problem for the conventional MISO-BC without the
PIPCs has also been studied in \cite{Boche04}, where an algorithm
was proposed using the virtual uplink formulation and a fixed-point
iteration. However, this algorithm cannot be extended directly to
deal with multiple PIPCs for the case of CR MISO-BC. Similarly as
for the previous discussions on the WSRMax problem for the CR
MIMO-BC where a generalized MIMO MAC-BC duality holds, a counterpart
beamforming duality also holds for the MISO-BC and SIMO-MAC
\cite{ZhangLan09a}. With this duality result, the SINR balancing
problem (P4) for the CR MISO-BC can be converted into an equivalent
problem for the dual SIMO-MAC, where the efficient iterative
algorithm in \cite{Boche04} can be directly applied. The interested
readers may refer to \cite{ZhangLan09a} for the details of this
method.

{\bf CR MIMO-IC}: Second, consider the CR MIMO-IC  given in
(\ref{eq:IC}), subject to both the PTPCs for the $K$ SU-TXs and the
PIPCs for the $J$ PUs, which can be similarly defined as for the MAC
case in (\ref{eq:PTPC}) and (\ref{eq:PIPC}), respectively. From an
information-theoretic perspective, the capacity region for the
Gaussian IC under PTPCs, which consists of all the simultaneously
achievable rates of all the users, still remains unknown in general
even for the case of $K=2$ and $A_k=B_k=1, k=1,2$ \cite{Han}. A
pragmatic approach that leads to suboptimal achievable rates in the
Gaussian IC is to restrict the system to operate in a decentralized
manner, i.e., allowing only single-user encoding and decoding by
treating the co-channel interferences from the other users as
additional Gaussian noises. For this approach, transmit optimization
for the CR MIMO-IC reduces to finding a set of optimal transmit
covariance matrices for the $K$ SU links, denoted by ${\bf
R}_k\in\mathbb{C}^{A_k\times A_k}, k=1,\cdots,K$, to maximize the
secondary network throughput under both the PTPCs and PIPCs. More
specifically, the WSRMax problem for the CR MIMO-IC can be expressed
as:
\begin{align}
\mathop{{\rm Max.}}_{{\bf R}_1,\cdots,{\bf R}_K} &
\sum_{k=1}^K\mu_k\log\bigg|{\bf I}+({\bf I}+\sum_{i\neq k}{\bf
H}_{ik}{\bf R}_i {\bf H}_{ik}^H)^{-1}{\bf H}_{kk}{\bf R}_k {\bf
H}_{kk}^H\bigg| \nonumber \\ &~~~~~~~~~~~~~~~~~~~~~~~~~~~~~~~~~~~~~~~~~~~~~~~~~~~~~{\rm(P5)} \nonumber  \\
{\rm s.~t.} & ~~{\rm Tr} ({\bf R}_k) \leq P_k, ~ k=1,\cdots,K
\nonumber
\\ & ~~\sum_{k=1}^{K}{\rm Tr} \left({\bf E}_{kj}{\bf R}_k{\bf
E}^H_{kj}\right) \leq \Gamma_j, ~ j=1,\cdots, J \nonumber
\\ & ~~{\bf R}_k\succeq {\bf 0}, ~ k=1,\cdots,K \nonumber
\end{align}
where $\mu_k$'s are the given non-negative user rate weights. We see
that (P5) is non-convex with or without the PIPCs due to the fact
that the objective function is non-concave over ${\bf R}_k$'s for
$K>1$. As a result, there are no efficient algorithms yet to obtain
the globally optimal solution for this problem. For the same problem
setup, there have been recent progresses on characterizing the
maximum achievable ``degrees of freedom (DoF)'' for the user
sum-rate (i.e., $\mu_k=1, \forall k$) \cite{Jafar08}.

Next, we discuss some feasible solutions for (P5). First, it is
worth noting that for (P5) in the case without the PIPCs, a commonly
adopted suboptimal approach is to iteratively optimize each user's
transmit covariance subject to its individual PTPC with the transmit
covariances of all the other users fixed. This decentralized
approach has been first proposed in \cite{Ingram,Blum03} to obtain
some local optimal points for (P5) with the PTPCs only, where they
differ in that the one in \cite{Ingram} maximizes the user
individual rate at each iteration, while the one in \cite{Blum03}
maximizes the user weighted sum-rate. It is also noted that a
parallel line of works with similar iterative user optimizations has
been pursued in the single-antenna but multi-carrier based
interference channels such as the wired discrete-multi-tone (DMT)
based digital subscriber line (DSL) network \cite{Yu02}, and the
wireless OFDM based ad hoc network \cite{Huang}. One important
question to answer for such iterative algorithms is under what
conditions the algorithm will guarantee to converge to a local
optimal point. This problem has been addressed in the contexts of
both multi-carrier and multi-antenna based interference channels in,
e.g., \cite{Luo06b,Palomar08a}, via game-theoretic approaches.

However, the above iterative approach cannot be applied directly to solve
(P5) with both the PIPCs and PTPCs, since each PIPC involves all the
user transmit covariances and is thus not separable over the SUs.
Thus, a feasible approach for (P5) is to decompose each of the $J$
PIPCs into a set of interference-power constraints over the $K$
SU-TXs, i.e., for the $j$th PIPC, $j\in\{1,\cdots,J\}$,
\begin{align}\label{eq:PIPC IC}
{\rm Tr}\left({\bf E}_{kj}{\bf R}_k{\bf E}_{kj}^H\right)\leq
\Gamma_{j}^{(k)}, ~ k=1,\cdots,K
\end{align}
where $\Gamma_{j}^{(k)}$ is a constant, and all
$\Gamma_{j}^{(k)}$'s, $k=1,\cdots,K$, satisfy
$\sum_k\Gamma_{j}^{(k)}\leq  \Gamma_{j}$ such that the $j$th PIPC is
guaranteed. Then, the iterative algorithm works here, where each SU
link independently optimizes ${\bf R}_k$ to maximize its achievable
rate under its PTPC and $J$ interference-power constraints given by
(\ref{eq:PIPC IC}), with all other ${\bf R}_i$'s, $i\neq k$, fixed.
It is observed that the resulting problem is in the same form of our
previously studied (P1) for the CR point-point MIMO channel; thus,
similar techniques developed for (P1) can be applied. Note that a
suboptimal method for this problem in the same spirit of the partial
channel projection method to reduce the design complexity for each
SU transmit covariance matrix has also been proposed in
\cite{Palomar08b}. Moreover, it is noted that $\Gamma_{j}^{(k)}$'s,
$j=1,\cdots,J$, $k=1,\cdots,K$, can be searched over the SUs to
further improve their weighted sum-rate.

Alternatively, assuming that a centralized optimization is feasible
with the global knowledge of all the channels in the SU network, as
well as those from different SU-TXs to all PUs, another heuristic
algorithm for (P5) was proposed in \cite{Kim08}. By rewriting the SU
transmit covariance matrices into their equivalent precoding vectors
and power allocation vectors, this algorithm iteratively updates the
SU transmit precoding vectors (based on the ``network duality''
\cite{NetworkDuality}) or the power allocation vectors (by solving
geometric programming (GP) problems \cite{Chiang07}), with the
others being fixed.

It is worth pointing out that there are other problem formulations
different from (P5) to address the transmit optimization for the CR
MIMO-IC. In \cite{Vu08}, a new criterion was proposed to design the
SU link transmission in a CR MISO-IC via an alternative
decentralized approach, where each SU-TX independently designs its
transmit precoding vector to maximize the ratio between the received
signal power at the desired SU-RX and the resulted total
interference power at all the PUs, in order to regulate the
interference powers at PUs. Moreover, the above discussions are all
based on the assumption that each SU-RX treats the interferences
from all the other SU links as additional noises, which is of
practical interest since it simplifies the receiver design for each
SU link. However, due to independent cross-link channels between SU
terminals, it may be possible that a SU-RX could occasionally
observe ``strong'' interference signals from some co-existing SU-TXs
and thus be able to decode their messages via multiuser detection
techniques and then cancel the associated interferences. With such
``opportunistic'' multiuser detection at each SU-RX, the achievable
rate of each SU link becomes a function of not only its own transmit
covariance, but also those of the other SUs as well as their
instantaneous transmit rates. Thus, the corresponding transmit
optimization for the CR MIMO-IC leads to new and more challenging
problem formulations than (P5); the interested readers may refer to
\cite{Tajer,ZhangISS1}.

\section{Joint Space-Time-Frequency DRA Optimization}
\label{sec:spacetime DRA}

In the previous section, we have studied DRA for different CR
networks at a single transmit dimension in time/frequency, by
focusing on spatial-domain transmit optimization under the {\it
peak} transmit and interference power constraints (PTPC and PIPC).
In this section, we bring the additional time and/or frequency
dimensions into the DRA problem formulations, by applying the {\it
average} transmit and interference power constraints (ATPC and AIPC)
in CR networks. Consider the DRA over $L$ time/freqeuncy dimensions,
for which all the required channel knowledge is assumed to be known.
Taking the CR MAC as an example (similar arguments can be developed
for the CR BC/IC), under both the ATPCs and AIPCs given in
(\ref{eq:ATPC}) and (\ref{eq:AIPC}), respectively, a generic problem
formulation for DRA optimization can be formulated as:
\begin{align}
\mathop{{\rm Max.}}_{{\bf S}_k[l]\succeq {\bf 0}, \forall k,l } & ~~
C(\{{\bf S}_k[l]\}) \nonumber ~~~~~~~~~~~~~~\rm{(P6)}
\\ {\rm s.~t.} & ~~  (\ref{eq:ATPC}), (\ref{eq:AIPC}) \nonumber
\end{align}
where $\{{\bf S}_k[l]\}$ denotes the set of ${\bf S}_k[l]$'s,
$k=1,\cdots,K$, and $l=1,\cdots,L$, while $C(\cdot)$ is an arbitrary
utility function to measure the CR network performance. We assume
that $C(\cdot)$ is separable over $l$'s, i.e., $C(\{{\bf
S}_k[l]\})=\frac{1}{L}\sum_{l=1}^L U_l({\bf S}_1[l],\cdots,{\bf
S}_K[l])$ with $U_l(\cdot)$'s denoting individual utility functions.
Since both the ATPC and AIPC involve $L$ transmit covariance
matrices, the {\it Lagrange dual decomposition} (see, e.g., a
tutorial paper \cite{Palomar06}) is a general method to deal with
this type of average constraints for optimization over a number of
parallel dimensions, which is explained as follows. By introducing a
set of dual variables, $\nu_k$'s, each for one of the $K$ ATPCs, and
$\delta_k$'s, each for one of the $J$ AIPCs, the Lagrange dual
problem of (P6) can be written as (P6-D):
\begin{align}
\mathop{{\rm Min.}}_{\mv{\nu}\succeq {\bf 0}, \mv{\delta}\succeq
{\bf 0}} d(\mv{\nu},\mv{\delta}) \nonumber
\end{align}
with $\mv{\nu}=[\nu_1,\cdots,\nu_K],
\mv{\delta}=[\delta_1,\cdots,\delta_J]$, and the dual function
\begin{align}\label{eq:Lagrangian}
d(\mv{\nu},\mv{\delta})\triangleq \max_{{\bf S}_k[l]\succeq {\bf 0},
\forall k,l } C(\{{\bf S}_k[l]\})-\sum_{k=1}^K
\nu_k(\frac{1}{L}\sum_{l=1}^L{\rm Tr} ({\bf S}_k[l]) \nonumber
\\ -\bar{P}_k)-\sum_{j=1}^J\delta_j(\frac{1}{L}\sum_{l=1}^L\sum_{k=1}^{K}{\rm Tr} \left({\bf
G}_{kj}[l]{\bf S}_k[l]{\bf G}^H_{kj}[l]\right) - \bar{\Gamma}_j).
\end{align}
Since the dual problem (P6-D) is convex regardless of the convexity
of the primal problem (P6) \cite{Boydbook}, (P6-D) can be
efficiently solved by the ellipsoid method according to the
subgradients of the dual function $d(\mv{\nu},\mv{\delta})$,
similarly as in our previous discussions, provided that the
maximization problem in (\ref{eq:Lagrangian}) is solvable for any
given set of $\mv{\nu}$ and $\mv{\delta}$. It is interesting to
observe that this maximization problem can be decomposed into $L$
parallel subproblems each for one of the $L$ dimensions, and all of
these subproblems have the same structure and are thus solvable by
the same algorithm, a practice known as ``dual decomposition''.
Without loss of generality, we drop the dimension index $l$ and
express each subproblem as
\begin{align}\label{eq:subdual}
\max_{{\bf S}_k\succeq {\bf 0}, \forall k} U({\bf S}_1,\cdots,{\bf
S}_K)-\sum_{k=1}^K {\rm Tr} \left({\bf B}_k(\nu_k,\mv{\delta}){\bf
S}_k\right)
\end{align}
where ${\bf B}_k(\nu_k,\mv{\delta})=\nu_k{\bf
I}+\sum_{j=1}^J(\delta_j{\bf G}_{kj}^H{\bf G}_{kj})$ is a constant
matrix for the given $\nu_k$ and $\mv{\delta}$, $k=1\cdots,K$.

We then discuss the following two cases. For the first case,
consider that $U_l(\cdot)$ is a concave function over ${\bf
S}_k[l]$'s, $\forall l$ (e.g., the point-to-point CR channel
capacity in (P1), or the weighted sum-rate for the CR MIMO-MAC in
(P2)). Then, (P6) is convex and thus the duality gap between the
optimal values of (P6) and (P6-D) is zero, i.e., (P6) and (P6-D) are
equivalent problems. Furthermore, each subproblem in
(\ref{eq:subdual}) is also convex. Thus, the dual decomposition
method solves (P6) via its dual problem (P6-D), which is
decomposable into $L$ convex subproblems. For the second case, as a
counterpart, consider that $U_l(\cdot)$ is non-concave over ${\bf
S}_k[l]$'s (e.g., the weighted sum-rate for the CR MIMO-BC/MIMO-IC
in (P3)/(P5)). As a result, (P6) is non-convex and the duality gap
between (P6) and (P6-D) may not be zero. Furthermore, the subproblem
(\ref{eq:subdual}) is also non-convex. For this case, even when the
optimal solutions of the $L$ subproblems are obtainable, the optimal
value of (P6-D) in general only serves as an upper bound on that of
(P6). However, in \cite{Yu06b} it is pointed out that if a set of
so-called ``time-sharing'' conditions are satisfied by a non-convex
optimization problem, the duality gap for this problem and its dual
problem is indeed zero. Furthermore, for the class of DRA problems
in the form of (P6), the associated time-sharing conditions are
usually satisfied asymptotically as $L\rightarrow \infty$ under some
cautious considerations on the continuity of channel distributions
\cite{Luo08}. Therefore, the dual decomposition method could still
be applied to solve (P6) in the non-convex case for sufficiently
large values of $L$, provided that the optimal solutions for the
subproblems in (\ref{eq:subdual}) are obtainable (e.g., a variation
of (P3) for the CR MIMO-BC). However, with finite values of $L$, how
to efficiently solve (P6) in the case of non-concave objective
functions is still open.

With the above discussions on the general approaches to design joint
space-time-frequency DRA for CR networks, we next present some
examples of unique interests to CR systems.

\subsection{TDMA/FDMA Constrained DRA: When Is It Optimal?}

Time-/frequency-division multiple-access (TDMA/FDMA), which
schedules only one user for transmission at each time/frequency
dimension, is usually preferable in practice due to their
implementation ease. For the TDMA/FDMA based CR MAC (similar
arguments hold for the CR BC/IC), the optimal DRA over $L$ transmit
dimensions to maximize the sum-capacity of the SUs can be formulated
as (P6) with properly chosen functions for $U_l(\cdot)$'s, where for
any given $l$, $U_l(\cdot)$ is expressed as ($l$ is dropped for
conciseness)
\begin{align}
U({\bf S}_1,\cdots,{\bf S}_K)=\left\{\begin{array}{ll}
\log\left|{\bf I}+{\bf H}_k{\bf S}_k{\bf H}_k^H\right| & ~ {\bf
S}_i={\bf 0}, \forall i\neq k \\ 0 & ~~ {\rm otherwise.}
 \end{array} \right.
\end{align}
Note that $U(\cdot)$ defined above implies the TDMA/FDMA constraint,
i.e., only scheduling one user for transmission at a given dimension
with a positive contribution to the sum-capacity. However, it can be
shown that $U(\cdot)$ is non-concave over ${\bf S}_k$'s  in this
case and as a result, the corresponding (P6) is non-convex.
Nevertheless, according to our previous discussions, since the
time-sharing conditions hold approximately when
$L\rightarrow\infty$, the dual decomposition method can be applied
to solve (P6) for this case with very large values of $L$, where the
optimal solution of the associated subproblem at each dimension
given in (\ref{eq:subdual}) can be obtained by finding the SU
(selected for transmission) with the largest objective value of the
following problem (which is of the same form as Problem (\ref{eq:max
P1}) and thus solvable in a similar way):
\begin{align}
\max_{{\bf S}_k\succeq {\bf 0}} ~ \log\left|{\bf I}+{\bf H}_k{\bf
S}_k{\bf H}_k^H\right| -{\rm Tr} ({\bf B}_k(\nu_k,\mv{\delta}){\bf
S}_k).
\end{align}

An important question to investigate for TDMA/FDMA based DRA is how
much the performance is degraded as compared with the optimal DRA
that allows more than one users to transmit at a given dimension.
From an information-theoretic viewpoint, it is thus pertinent to
investigate the conditions for the optimality of TDMA/FDMA, i.e.,
when they are optimal to achieve the system sum-capacity. For the
traditional single-antenna fading MAC under the user ATPCs over
time, it has been shown in \cite{Knopp95} that TDMA is optimal for
achieving the ergodic/long-term sum-capacity. This result has been
shown to hold for the fading CR MAC and CR BC under both the ATPCs
and AIPCs in \cite{Zhang09MAC}, where by exploiting the KKT
optimality conditions of the associated optimization problems, the
optimality conditions for TDMA in other cases of combined
peak/average transmit/interference power constraints have been
characterized. For the traditional single-antenna IC with
interference treated as noise, the optimality of TDMA/FDMA for the
sum capacity has been investigated under the ATPCs in
\cite{Tse07,Luo}. It would be interesting to extend these results to
the case of CR IC under the additional PIPCs and/or AIPCs.

\subsection{Peak vs. Average Interference Power Constraints: A New Interference Diversity}

From a SU's perspective, it is obvious that the ATPC/AIPC is more
flexible than the PTPC/PIPC for DRA under the same power threshold
and thus results in a larger SU link capacity. However, from a PU's
perspective, it remains unknown whether the AIPC or PIPC causes more
PU link performance degradation. Intuitively speaking, the PIPC
should be more favorable than the AIPC since the former limits the
interference power at the PU to be below certain threshold at each
time/frequency dimension, while the latter results in variations of
interference power levels over different dimensions although their
average level is kept below the same threshold as that for the PIPC.

Somehow surprisingly, in \cite{Zhang09} it is shown that for the
single-antenna PU fading channel subject to the interference from a
SU transmitter, the AIPC is in fact better than its PIPC counterpart
under the same average power threshold in terms of minimizing the PU
capacity losses, which holds for the cases of both ergodic and
outage capacities of the PU channel, with/without power control. To
illustrate this result, we consider for simplicity the case without
the PU link power control, i.e., the PU transmits with a constant
power, $Q$, over all the fading states. Suppose that the PU link
channel power gain is denoted by $h_p$, and that from the SU
transmitter to the PU receiver denoted by $h_{sp}$. Next, consider
the following two cases, where the interference power from the SU
transmitter at the PU receiver, denoted by $I_{sp}=h_{sp}p_s$, with
$p_s$ denoting the SU transmit power, is fixed over all the fading
states in Case I (corresponding to the case of PIPC), and is allowed
to be variable in Case II (corresponding to the case of AIPC). For
both cases, a constant interference power threshold $\Gamma$ is set
and is assumed to hold with equality, i.e., for Case I, $I_{sp}^{\rm
(I)}=\Gamma$, for all the fading states, while for Case II, ${\rm
E}(I_{sp}^{\rm (II)})=\Gamma$. Taking the PU ergodic capacity as an
example, which can be expressed as (assuming unit-power receiver
Gaussian noise):
\begin{align}
C_p={\rm E}\left(\log\left(1+\frac{h_pQ}{1+I_{sp}}\right)\right).
\end{align}
Let $C_p^{\rm (I)}$ and $C_p^{\rm (II)}$ denote the values of $C_p$
in Cases I and II, respectively. The following
equalites/inequalities then hold
\begin{align}
C_p^{\rm (I)}&= {\rm
E}_{h_p}\left(\log\left(1+\frac{h_pQ}{1+\Gamma}\right)\right) \nonumber \\
&={\rm E}_{h_p}\left(\log\left(1+\frac{h_pQ}{1+{\rm
E}(I_{sp}^{\rm(II)})}\right)\right) \nonumber \\
&\leq {\rm E}_{h_p}\left({\rm
E}_{I_{sp}}\left(\log\left(1+\frac{h_pQ}{1+I_{sp}^{\rm
(II)}}\right)\right)\right) \label{eq:1} \\ &=C_p^{\rm (II)}
\nonumber
\end{align}
where (\ref{eq:1}) is due to the Jensen's inequality (see, e.g.,
\cite{Cover}) and the convexity of the function
$f(x)=\log\left(1+\frac{\kappa}{1+x}\right)$ where $\kappa$ is any
positive constant and $x\geq 0$. Thus, it follows that given the
same average power of the interference, $\Gamma$, it is desirable
for the PU to have the instantaneous interference power $I_{sp}$
fluctuate over fading states (Case II) rather than stay constant
(Case I), to achieve a larger ergodic capacity.

In general, the results in \cite{Zhang09}  reveal a new {\it
interference diversity} phenomenon for SS-based CR networks, i.e.,
the randomized interference powers from the secondary network can be
more advantageous over deterministic ones across different transmit
dimensions over space, time, or frequency for minimizing the
resulted primary network capacity losses. Further investigations are
required on interference diversity driven DRA for CR or other
spectrum sharing systems.

\subsection{Beyond Interference Temperature: Exploiting Primary Link Performance Margins}

So far, we have studied DRA for CR networks based on the IT
constraints for protecting the PU transmissions. Given that the IT
constraints in general conservatively lead to an upper bound on the
PU capacity loss due to the interference from the SUs
\cite{Zhang08a,Zhang08b}, it would be possible to improve the
spectrum sharing capacities for both the SUs and PUs over the
IT-based methods if additional cognition on the PU transmissions is
available at the CR transmitters. For example, by exploiting CSI of
the PU links, the CRs could allocate transmit/interference powers
more flexibly over the dimensions where the PU channels exhibit poor
conditions, without degrading too much the PU link performances.
These PU ``null'' dimensions could come up in time, frequency, or
space. Thus, the IT constraints could be replaced by the more
relevant {\it primary link performance margin constraints}
\cite{Zhang08b,Kang09b} for the design of DRA in CR networks, in
order to optimally exploit the available primary link performance
margins to accommodate the interference from the SUs. Following this
new paradigm, many new and challenging DRA problems can be
formulated for CR networks. As an example, consider the same setup
with a pair of single-antenna PU and SU links over fading channels
as in the previous subsection. Instead of applying the conventional
AIPC: ${\rm E}(h_{sp}p_s)\leq\Gamma$, over the SU power allocation,
we may apply the following PU ergodic capacity constraint
\cite{Zhang08b}
\begin{align}\label{eq:PU loss constraint}
{\rm E}\left(1+\frac{h_pQ}{1+h_{sp}p_s}\right)\geq \bar{C}_p
\end{align}
where $\bar{C}_p$ is a given threshold for the minimum PU ergodic
capacity. Note that the new constraint in (\ref{eq:PU loss
constraint}) is more directly related to the PU transmission than
the conventional AIPC. However, it can be verified that the
constraint in (\ref{eq:PU loss constraint}) is non-convex over $p_s$
in general, thus resulting in more challenging SU power allocation
problems than that with the convex AIPC. The optimal power
allocation rules for the SU link subject to the AIPC vs. the newly
introduced PU ergodic capacity constraint given by (\ref{eq:PU loss
constraint}) are compared in \cite{Zhang08b}, where it is shown that
the new constraint achieves notable rate improvements for both the
PU and SU links over the conventional AIPC.

\section{Conclusions and Directions for Future Work}\label{sec:conclusion}

Dynamic resource allocation (DRA) has become an essential building
block in CR networks to exploit various cognitions over both the
primary and secondary networks for CR transmit optimization subject
to certain required primary protection. In this article, we have
presented an extensive list of new, challenging, and unique problems
for designing the optimal DRA in CR networks, and demonstrated the
key role of various convex optimization techniques in solving the
associated design problems. In addition to those open issues as
highlighted in our previous discussions, other promising areas of
practical and theoretical interests are discussed as follows, which
open an avenue for future work.

{\it Robust Cognitive Beamforming}: In our previous discussions on
cognitive beamforming, we have observed that the knowledge of
channels from each secondary transmitting terminal to all PUs is
essential to the design optimization. However, since the primary and
secondary networks usually belong to different operators, it is
difficult for the PUs to feed back the required CSI to the CRs. As a
result, the SU usually needs to rely on its own observations over
the received signals from the primary terminals to extract the
required CSI \cite{Zhang08c}. Nevertheless, the estimated CSI on the
SU-to-PU channels may contain errors, which should be taken into
account for the design of practical CR systems. This motivates a new
and challenging research direction on robust designs for cognitive
beamforming to cope with imperfect CSI \cite{ZhangLan09b,Zheng09}.
More investigations on the robust cognitive beamforming designs for
more general CR networks and CSI uncertainty models are appealing.

{\it Active Interference-Temperature Control}: In this article, we
have focused on the design of CR networks subject to the given
interference-power constraints for protecting the PUs. We have also
discussed some promising rules on how to optimally set the IT
constraints in the CR network to achieve the best spectrum sharing
throughput. These results lead to a new and universal design
paradigm for interference management in CR or other related
multiuser communication systems \cite{ZhangLanSecrecy,Zhang10}, via
appropriately setting the IT levels across the coexisting links. The
active IT control approach to interference management for multiuser
communication systems is relatively new, and more research endeavors
are required along this direction.

\vspace{0.1in}

\hspace{-0.2in} {\bf AUTHORS:}

{\it Rui Zhang} (rzhang@i2r.a-star.edu.sg) received the B.Eng. and
M.Eng. degrees in electrical and computer engineering from National
University of Singapore in 2000 and 2001, respectively, and the
Ph.D. degree in electrical engineering from Stanford University,
California, USA, in 2007. He is now a Senior Research Fellow with
the Institute for Infocomm Research (I2R), Singapore. He also holds
an Assistant Professorship position in electrical and computer
engineering with National University of Singapore. He has
authored/co-authored more than 100 refereed international journal
and conference papers. His current research interests include
cognitive radios, cooperative communications, and multiuser MIMO
systems.

{\it Ying-Chang Liang} (ycliang@i2r.a-star.edu.sg) is presently a
Senior Scientist in the Institute for Infocomm Research (I2R),
Singapore. He also holds adjunct associate professorship positions
in Nanyang Technological University (NTU) and National University of
Singapore (NUS). He has been teaching graduate courses in NUS since
2004 and was a visiting scholar with the Department of Electrical
Engineering, Stanford University, CA, USA, from Dec 2002 to Dec
2003. His research interest includes cognitive radio, dynamic
spectrum access, reconfigurable signal processing for broadband
communications, space-time wireless communications, wireless
networking, information theory and statistical signal processing.

{\it Shuguang Cui} (cui@tamu.edu) received Ph.D in Electrical
Engineering from Stanford University in 2005, M.Eng in Electrical
Engineering from McMaster University, Canada, in 2000, and B.Eng. in
Radio Engineering with the highest distinction from Beijing
University of Posts and Telecommunications, China, in 1997. He is
now working as an assistant professor in Electrical and Computer
Engineering at the Texas A\&M University, College Station, TX. His
current research interests include cross-layer optimization for
resource-constrained networks and network information theory. He has
been serving as the associate editors for the IEEE Communication
Letters and IEEE Transactions on Vehicular Technology, and the
elected member for IEEE Signal Processing Society SPCOM Technical
Committee.


\begin{thebibliography}{1}
\bibliographystyle{IEEEbib}

%%%%%%%%%%%Survey%%%%%%%%%%%%%%%%%%

\bibitem{Mitola} J. Mitola and G. Q. Maguire, ``Cognitive radio: making
software radios more personal,'' {\it IEEE Pers. Commun.}, vol. 6,
no. 4, pp. 13-18, Aug. 1999.

\bibitem{Zhao07} Q. Zhao and B. M. Sadler, ``A survey of dynamic spectrum
access,'' {\it IEEE Sig. Proces. Mag.}, vol. 24, no. 3, pp. 79-89,
May 2007.

\bibitem{Goldsmith08} A. Goldsmith, S. A. Jafar, I.
Mari\'{c}, and S. Srinivasay, ``Breaking spectrum gridlock with
cognitive radios: an information theoretic perspective,'' {\it Proc.
IEEE}, vol. 97, no. 5, pp. 894-914, May, 2009.

\bibitem{Quan08a} Z. Quan, S. Cui, and A. Sayed, ``Optimal linear cooperation for
spectrum sensing in cognitive radio networks,'' {\it IEEE J. Sel.
Topics Sig. Proces.}, vol. 2, no. 1, pp. 28-40, Feb. 2008.

\bibitem{Liang08} Y.-C. Liang, Y. Zeng, E. C. Y. Peh, and A. T. Hoang,
``Sensing-throughput tradeoff for cognitive radio networks,'' {\it
IEEE Trans. Wireless Commun.}, vol. 7, no. 4, pp. 1326-1337, Apr.
2008.

\bibitem{Li08} B. H. Juang, Y. Li, and J. Ma, ``Signal processing in cognitive
radio,'' {\it Proc. IEEE}, vol. 97, no. 5, pp. 805-823, May 2009.

\bibitem{Zeng} Y. H. Zeng, Y.-C. Liang, A. T. Hoang, and R. Zhang, ``A review on
spectrum sensing for cognitive radio: challenges and solutions,''
{\it EURASIP J. Advances in Sig. Proces.}, Article ID 381465, 2010.

\bibitem{Kumar00} P. Gupta and P. R. Kumar, ``The capacity of wireless networks,''
{\it IEEE Trans. Inf. Theory}, vol. 46, no. 3, pp. 388-404, Mar.
2000.

\bibitem{Tarokh06} N. Devroye, P. Mitran, and V. Tarokh, ``Achievable
rates in cognitive radio channels,'' {\it IEEE Trans. Inf. Theory},
vol. 52, no. 5, pp. 1813-1827, May 2006.

\bibitem{Gastpar07} M. Gastpar, ``On capacity under receive and spatial spectrum-sharing
constraints,'' {\it IEEE Trans. Inf. Theory}, vol.~53, no.~2, pp.
471-487, Feb. 2007.

\bibitem{Luo06a} Z.-Q. Luo and W. Yu, ``An introduction to convex optimization for
communications and signal processing,'' {\it IEEE J. Sel. Areas
Commun.}, vol. 24, no. 8, pp. 1426-1438, Aug. 2006.

\bibitem{Shamai98} E. Biglieri, J. Proakis, and S. Shamai (Shitz), ``Fading
channels: information-theoretic and communications aspects,'' {\it
IEEE Trans. Inf. Theory}, vol. 44, no. 6, pp. 2619-2692, Oct. 1998.

%%%%%%%%%%%%%%%%%MIMO%%%%%%%%%%%%%%%%%%%%%%%%

\bibitem{Zhang08a} R. Zhang and Y.-C. Liang, ``Exploiting multi-antennas for
opportunistic spectrum sharing in cognitive radio networks,'' {\it
IEEE J. S. Topics Sig. Proces.}, vol. 2, no. 1, pp. 88-102, Feb.
2008.

\bibitem{Boydbook} S. Boyd and L. Vandenberghe, {\it Convex Optimization}, Cambridge University Press, 2004.

\bibitem{CVX} M. Grant and S. Boyd, ``CVX: Matlab software for disciplined convex
programming,'' available [online] at http://stanford.edu/~boyd/cvx.

\bibitem{BGT81} R. G. Bland, D. Goldfarb, and M. J. Todd, ``The ellipsoid method:
a survey,'' {\it Operations Research,} vol. 29, no. 6, pp.
1039-1091, 1981.

\bibitem{Cover} T. Cover and J. Thomas, {\it Elements of information theory,} New York: Wiley, 1991.

\bibitem{Spencer04} Q. H. Spencer, A. L. Swindlehurst, and M. Haardt, ``Zero-forcing
methods for downlink spatial multiplexing in multiuser MIMO
channels,'' {\it IEEE Trans. Sig. Proces.}, vol. 52, no. 2, pp.
461-471, Feb. 2004.

\bibitem{Tse98} D. Tse and  S. Hanly,``Multi-access fading channels-Part I: polymatroid structure, optimal resource
allocation and throughput capacities,'' {\it IEEE Trans. Inf.
Theory}, vol. 44, no. 7, pp. 2796-2815, Nov. 1998.


\bibitem{ZhangLan08a} L. Zhang, Y.-C. Liang, and Y. Xin, ``Joint beamforming and power
control for multiple access channels in cognitive radio networks,''
{\it IEEE J. Sel. Areas Commun.}, vol.26, no.1, pp.38-51, Jan. 2008.

%%%%%%%%%%%%%%BC%%%%%%%%%%%%%%%%%%%%

\bibitem{multicasting} K. Phan, S. Vorobyov, N. Sidiropoulos, and C. Tellambura,
``Spectrum sharing in wireless networks via QoS-aware secondary
multicast beamforming,'' {\it IEEE Trans. Sig. Proces.}, vol. 57,
no. 6, pp. 2323-2335, June 2009.

\bibitem{Shamai06} H. Weingarten, Y. Steinberg, and S. Shamai (Shitz), ``The capacity region of the Gaussian
multiple-input multiple-output broadcast channel,'' {\it IEEE Trans.
Inf. Theory}, vol. 52, no. 9, pp. 3936-3964, Sep. 2006.

\bibitem{Goldsmith03} S. Vishwanath, N. Jindal, and A. Goldsmith, ``Duality, achievable rates, and sum-rate capacity of
Gaussian MIMO broadcast channels,'' {\it IEEE Trans. Inf. Theory,}
vol. 49, no. 10, pp. 2658-2668, Oct. 2003.

\bibitem{Yu} W. Yu and T. Lan, ``Transmitter optimization for the multi-antenna
downlink with per-antenna power constraints,'' {\it IEEE Trans. Sig.
Proces.}, vol. 55, no. 6, pp. 2646-2660, June 2007.

\bibitem{ZhangLan09a} L. Zhang, R. Zhang, Y.-C. Liang, Y. Xin, and H. V. Poor, ``On the
Gaussian MIMO BC-MAC duality with multiple transmit covariance
constraints,'' submitted to {\it IEEE Trans. Inf. Theory}. Available
[online] at arXiv:0809.4101.

\bibitem{Ottersten01}M. Bengtsson and B. Ottersten, ``Optimal and suboptimal transmit
beamforming,'' in {\it Handbook of Antennas in Wireless
Communications}, CRC Press, 2001.

\bibitem{Wiesel06} A. Wiesel, Y. C. Eldar, and S. Shamai (Shitz), ``Linear precoding
via conic optimization for fixed MIMO receivers,'' {\it IEEE Trans.
Sig. Process.}, vol. 54, no. 1, pp. 161-176, Jan. 2006.

\bibitem{Boche04} M. Schubert and H. Boche, ``Solution of the multiuser
downlink beamforming problem with individual SINR constraints,''
{\it IEEE Trans. Veh. Technol.}, vol. 53, no. 1, pp. 18-28, Jan.
2004.

%%%%%%%%%%%%IC%%%%%%%%%%%%%%%%%%%%%

\bibitem{Han} T. S. Han and K. Kobayashi, ``A new achievable rate region for
the interference channel,'' {\it IEEE Trans. Inf. Theory}, vol. 27,
no. 1, pp. 49-60, Jan. 1981.

\bibitem{Jafar08} V. R. Cadambe and S. A. Jafar, ``Interference alignment and
the degrees of freedom for the K user interference channel,'' {\it
IEEE Trans. Inf. Theory}, vol. 54, no. 8, pp. 3425-3441, Aug. 2008.

\bibitem{Ingram} M. F. Demirkol and M. A. Ingram, ``Power-controlled capacity for
interfering MIMO links,'' in {\it Proc. IEEE Vehicular Tech. Conf.
(VTC)}, vol. 1 pp. 187-191, 2001.

\bibitem{Blum03} S. Ye and R. S. Blum, ``Optimized signaling for MIMO interference
systems with feedback,'' {\it IEEE Trans. Sig. Process.}, vol. 51,
pp. 2839-2848, Nov. 2003.


\bibitem{Yu02} W. Yu, G. Ginis, and J. Cioffi ``Distributed multiuser power
control for digital subscriber lines,'' {\it IEEE J. Sel. Areas
Commun.}, vol. 20, no.5, pp. 1105-1115, June 2002.

\bibitem{Huang} J. Huang, R. Berry, and M. L. Honig, ``Distributed interference
compensation in wireless networks,'' {\it IEEE J. Sel. Areas
Commun.}, vol. 24, no. 5, pp. 1074-1084, May 2006.

\bibitem{Luo06b} Z.-Q. Luo and J.-S. Pang, ``Analysis of iterative waterfilling
algorithm for multiuser power control in digital subscriber line,''
{\it EURASIP J. Appl. Sig. Process.}, Article ID 24012, 2006.

\bibitem{Palomar08a} G. Scutari, D. P. Palomar, and S. Barbarossa, ``Competitive
design of multiuser MIMO systems based on game theory: a unified
view,'' {\it IEEE J. Sel. Areas Commun.} vol. 25, no. 7, pp.
1089-1103, Sep. 2008.

\bibitem{Palomar08b} G. Scutari, D. P. Palomar, and S. Barbarossa,
``Cognitive MIMO radio,'' {\it IEEE Sig. Proces. Mag.}, vol. 25, no.
6, Nov. 2008.

\bibitem{Kim08} S. J. Kim and G. B. Giannakis, ``Optimal resource allocation for MIMO ad hoc cognitive radio
networks,'' in {\it Proc. Annual Allerton Conf. Commun. Control
Comput.}, pp. 39-45, Sep. 2008.

\bibitem{NetworkDuality} B. Song, R. Cruz, and B. Rao, ``Network duality for multiuser
mimo beamforming networks and applications,'' {\it IEEE Trans.
Commun.}, vol. 55, no. 3, pp. 618-630, Mar. 2007.

\bibitem{Chiang07} M. Chiang, C. W. Tan, D. P. Palomar, D. Neill, and D. Julian,
``Power control by geometric programming,'' {\it IEEE Trans.
Wireless Commun.}, vol. 6, no. 7, pp. 2640-2651, July 2007.

\bibitem{Vu08} S. Yiu, M. Vu, and V. Tarokh, ``Interference reduction by
beamforming in cognitive networks,'' in {\it Proc. IEEE Global
Commun. Conf. (GLOBECOM)}, pp. 1-6, Dec. 2008.

\bibitem{Tajer} A. Tajer, N. Prasad, and X. Wang, ``Beamforming and  rate
allocation in  MISO cognitive radio networks,'' {\it IEEE Trans.
Sig. Proces.}, vol. 58, no. 1, pp. 362-377, Jan. 2010.

\bibitem{ZhangISS1} R. Zhang and J. Cioffi, ``Iterative spectrum shaping with
opportunistic multiuser detection,'' in {\it Proc. IEEE Int. Symp.
Inf. Theory (ISIT)}, June 2009.

%%%%%%%%%%%%%%%Dual Decomposition%%%%%%%%%%%%%%%%%%%%%%%%%%%%%%%%

\bibitem{Palomar06} D. P. Palomar and M. Chiang, ``A tutorial on decomposition methods for
network utility maximization,'' {\it IEEE J. Sel. Areas Commun.},
vol. 24, no. 8, pp. 1439-1451, Aug. 2006.

\bibitem{Yu06b} W. Yu and R. Lui, ``Dual methods for nonconvex spectrum optimization of multicarrier
systems,'' {\it IEEE Trans. Commun.}, vol. 54. no. 7. pp. 1310-1322,
July 2006.

\bibitem{Luo08} Z.-Q. Luo and S. Zhang, ``Dynamic spectrum management: complexity
and duality,'' {\it IEEE J. S. Topics Sig. Proces.}, vol. 2, no. 1,
pp. 57-73, Feb. 2008.

%%%%%%%%%%multiuser diversity%%%%%%%%%%%%%%%%%%%

\bibitem{Knopp95} R. Knopp and P. A. Humblet, ``Information capacity and power control in single-cell multi-user
communications,'' in {\it Proc. IEEE Int. Conf. Comm. (ICC)}, pp.
331-335, 1995.

\bibitem{Zhang09MAC} R. Zhang, S. Cui, and Y.-C. Liang, ``On ergodic
sum capacity of fading cognitive multiple-access and broadcast
channels,'' {\it IEEE Trans. Inf. Theory}, vol. 55, no. 11, pp.
5161-5178, Nov. 2009.

\bibitem{Tse07} R. Etkin, A. Parekh, and D. Tse, ``Spectrum sharing for unlicensed
bands,'' {\it IEEE J. Sel. Areas in Commun.}, vol. 25, no. 3, pp.
517- 528, Apr. 2007.

\bibitem{Luo} S. Hayashi and Z.-Q. Luo, ``Spectrum management for
interference-limited multiuser communication systems,'' {\it IEEE
Trans. Inf. Theory}, vol. 55, no. 3, pp. 1153-1175, Mar. 2009.

%%%%%%%%%%Fading%%%%%%%%%%%%%%%%%%%


\bibitem{Zhang09} R. Zhang, ``On peak versus average interference
power constraints for protecting primary users in cognitive radio
networks,'' {\it IEEE Trans. Wireless Commun.}, vol. 8, no. 4, pp.
2112-2120, Apr. 2009.


%%%%%%%%%%%%%%%%%%Primary Link Margins%%%%%%%%%%%%%%%%%%%%%%%%%%%%%%%
\bibitem{Zhang08b} R. Zhang, ``Optimal power control over fading cognitive radio
channels by exploiting primary user CSI,'' in {\it Proc. IEEE Global
Commun. Conf. (GLOBECOM)}, Nov. 2008.

\bibitem{Kang09b} X. Kang, R. Zhang, Y.-C. Liang, and H. K. Garg, ``Optimal power
allocation for cognitive radio under primary user outage capacity
constraint,'' in {\it Proc. IEEE Int. Conf. Commun. (ICC)}, June
2009.

%%%%%%%%%%%Robust Beamforming%%%%%%%%%%%%%%%

\bibitem{Zhang08c} R. Zhang, F. Gao, and Y.-C. Liang, ``Cognitive beamforming made practical: effective
interference channel and learning-throughput tradeoff,'' to appear
in {\it IEEE Trans. Commun.}.

\bibitem{ZhangLan09b} L. Zhang, Y.-C. Liang, Y. Xin, and H. V. Poor,
``Robust cognitive beamforming with partial channel state
information,'' {\it IEEE Trans. Wireless Commun.}, vol. 8, no. 8,
pp. 4143-4153, Aug. 2009.

\bibitem{Zheng09} G. Zheng, K.-K.
Wong, and B. Ottersten, ``Robust cognitive beamforming with bounded
channel uncertainties,'' to appear in {\it IEEE Trans. Sig.
Proces.}.

%%%%%%%%%%%%%Extensions%%%%%%%%%%%%%%%%%%%%%%

\bibitem{ZhangLanSecrecy} L. Zhang, R. Zhang, Y.-C. Liang, Y. Xin, and S. Cui, ``On the
relationship between the multi-antenna secrecy communications and
cognitive radio communications,''  in {\it Proc. Annual Allerton
Conf. Commun. Control and Comput.}, 2009.

\bibitem{Zhang10} R. Zhang and S. Cui, ``Cooperative interference management in
multi-cell downlink beamforming,'' submitted to {\it IEEE Trans.
Sig. Proces.}. Available [online] at arXiv: 0910.2771.

\end{thebibliography}
\end{document}